\documentclass[10pt, pre, aps, twocolumn, superscriptaddress, longbibliography]{revtex4-2}

\usepackage{amsmath}
\usepackage{amssymb}
\usepackage{physics}
\usepackage{nicefrac}
\usepackage{orcidlink}
\usepackage{xcolor}
\definecolor{monbleu}{RGB}{76,114,176}
\usepackage{graphicx}
\usepackage[T1]{fontenc}
\usepackage[utf8]{inputenc}
\hypersetup{colorlinks=true, allcolors=monbleu}

\newcommand{\ga}{{\ensuremath{\{\!+\!\}}}}
\newcommand{\gp}{{\ensuremath{\{\!-\!\}}}}
\newcommand{\gap}{{\ensuremath{\{\!\pm\!\}}}}
\newcommand{\s}{{\ensuremath{\mathrm{s}}}}

\begin{document}

\title{The impact of behavioral homophily and conformity on epidemic spreading in networks with large groups}

\author{Olivier \surname{Ribordy}\orcidlink{0009-0009-9976-7714}}%
\affiliation{D\'epartement de physique, de g\'enie physique et d'optique, Universit\'e Laval, Qu\'ebec (Qc), Canada}%
\affiliation{Centre interdisciplinaire en mod\'elisation math\'ematique de l'Universit\'e Laval, Qu\'ebec (Qc), Canada}%
\author{Clara \surname{Granell}\orcidlink{0000-0003-3156-0417}}
\affiliation{Departament d'Enginyeria Inform\`{a}tica i Matem\`{a}tiques, Universitat Rovira i Virgili, Tarragona, Spain}
\affiliation{ComSCIAM, Universitat Rovira i Virgili, 43007 Tarragona, Spain}

\author{Guillaume \surname{St-Onge}\orcidlink{0000-0001-5415-3748}}
\affiliation{Department of Physics, the Roux Institute at Northeastern University, Portland, ME, USA}
\affiliation{Network Science Institute, the Roux Institute at Northeastern University, Portland, ME, USA}

\author{Laurent \surname{H\'ebert-Dufresne}\orcidlink{0000-0002-0008-3673}}
\affiliation{Vermont Complex Systems Institute, University of Vermont, Burlington, VT, USA}
\affiliation{Department of Computer Science, University of Vermont, Burlington, VT, USA}
\affiliation{Santa Fe Institute, Santa Fe, New Mexico, USA}

\author{Alex \surname{Arenas}\orcidlink{0000-0003-0937-0334}}%
\affiliation{Departament d'Enginyeria Inform\`{a}tica i Matem\`{a}tiques, Universitat Rovira i Virgili, Tarragona, Spain}
\affiliation{ComSCIAM, Universitat Rovira i Virgili, 43007 Tarragona, Spain}
\affiliation{Complexity Science Hub Vienna, 1030 Vienna, Austria}
\affiliation{Pacific Northwest National Laboratory, 902 Battelle Boulevard, Richland, Washington 99354, USA}%

\author{Antoine \surname{Allard}\orcidlink{0000-0002-8208-9920}}%
\affiliation{D\'epartement de physique, de g\'enie physique et d'optique, Universit\'e Laval, Qu\'ebec (Qc), Canada}%
\affiliation{Centre interdisciplinaire en mod\'elisation math\'ematique de l'Universit\'e Laval, Qu\'ebec (Qc), Canada}%
\affiliation{Vermont Complex Systems Institute, University of Vermont, Burlington, VT, USA}

\date{\today}

\begin{abstract}
    Understanding how social behavior influences epidemic dynamics has become a central focus in mathematical epidemiology.
    In particular, \textit{behavioral homophily} (the tendency of individuals to associate with similar others) and \textit{conformity} (the adjustment of individual behavior to group norms) are key mechanisms in shaping transmission patterns.
    In this work, we investigate the combined impact of these behavioral processes on the susceptible-infected-susceptible (SIS) dynamics on networks with large, densely connected groups, modeled as cliques.
    Each individual has an intrinsic behavioral preference, but their expressed behavior within a group is modulated by its composition, reflecting conformity dynamics.
    Using the approximate master equations (AME) framework, we characterize the interplay between behavioral heterogeneity, group structure, and epidemic localization.
    Our results reveal that behavioral homophily amplifies the effects of conformity in large groups, enabling minority behaviors to persist as well as substantially shifting epidemic thresholds and spreading regimes.
\end{abstract}

\maketitle

\begin{table*}[t]\caption{Table of Notation}
\begin{center}
\begin{tabular}{r c p{14cm} }
\multicolumn{3}{c}{\underline{Structural parameters}}\\
\multicolumn{3}{c}{}\\
$m$ & : & Membership, the number of groups an individual belongs to. $m\in \{m_{\mbox{min}}, \dots, m_{\mbox{max}}\}$\\
$n$ & : & Size of a group, the number of individuals belonging to the group. $n\in \{n_{\mbox{min}}, \dots, n_{\mbox{max}}\}$\\
$n^+$ & : & Number of active individuals in a group\\
$g_m$ & : & Distribution of individual memberships\\
$p_n$ & : & Distribution of group sizes\\
$\gamma_m, \gamma_n$ & : & Exponents of the power-laws used for the membership and group size distributions\\
$A_+, A_-$ & : & Fraction of active (resp. passive) individuals in the population, $A_- = 1-A_+$.\\
$h$ & : & Homophily parameter for mixing between active and passive types. $h \in [0, 1]$, with $h=0$ as random mixing and $h=1$ as full homophily.\\
$P_{n^+|n}$ & : & Composition distribution, the probability that a group of size $n$ contains $n^+$ active individuals.\\
$P_{\ga|n, n^+}$ & : & Response probability, the probability that a group of size $n$ containing $n^+$ active nodes adopts the active behavior.\\
$P_{\ga|n}$ & : & Fraction of groups of size $n$ that adopt active behavior.\\
$P_{\ga|n, \pm}$ & : & Probability that a group of size $n$ reached through an individual of type $\pm$ adopts active behavior.\\
$\sigma$ & : & Steepness of the response probability.\\
$\tau$ & : & Influence threshold, value of $n^+/n$ at which the response probability is exactly $0.5$.\\
$A_{+ | n, \gap}$ & : & Probability that an individual drawn at random from a group of size $n$ and activity $\gap$ is of the active type.\\
$A_{+ |\s, n, \gap}$ & : & Probability that a susceptible individual drawn at random from a group of size $n$ and activity $\gap$ is of the active type.\\
$\Omega(g_m, p_n)$ & : & Coupling between groups, corresponding to the average number of external neighbors for an individual chosen at random in a group.\\
$w^\gap$ & : & Expected fraction of the edges of a passive individual that are found in groups of type $\gap$.\\
\multicolumn{3}{c}{}\\
\multicolumn{3}{c}{\underline{Epidemiological parameters}}\\
\multicolumn{3}{c}{}\\
$\beta^\gp$, $\beta$ & : & Transmission rate of the disease in an unprotected interaction.\\
$\beta^\ga$ & : & Transmission rate of the disease in a protected interaction.\\
$\beta_c$ & : & Critical value of the unprotected transmission rate above which the endemic state is stable.\\
$\epsilon$ & : & Efficiency of the NPIs at preventing transmission. $\epsilon \in [0, 1]$, such that $\beta^\ga = \epsilon\beta^\gp$.\\
$\mu$ & : & Recovery rate. Time is rescaled such that $\mu = 1$ without loss of generality.\\
\multicolumn{3}{c}{}\\
\multicolumn{3}{c}{\underline{Dynamical variables}}\\
\multicolumn{3}{c}{}\\
$v_{\s|m, \pm}(t)$ & : & Probability that an individual of membership $m$ and type $\pm$ is susceptible at time $t$.\\
$v_{\s|m}(t)$ & : & Probability that an individual of membership $m$ is susceptible at time $t$ without regard to type.\\
$v_{\s|\pm}(t)$ & : & Probability that an individual of type $\pm$ is susceptible at time $t$ without regard to membership.\\
$c_{i|n, \gap}(t)$ & : & Probability that a group of size $n$ and activity $\gap$ contains $i$ infected individuals at time $t$.\\
$c_{i|n}(t)$ & : & Probability that a group of size $n$contains $i$ infected individuals at time $t$, without regard to activity.\\
$r^\pm(t)$ & : & Average infection rate received from a group to which the individual of type $\pm$ belongs.\\
$\rho_n^\gap(t)$ & : & Average infection rate received from all \textit{other} groups by an individual chosen at random in a group of size $n$ and activity $\gap$.\\
$I_n^\gap(t)$ & : & Prevalence of the disease within groups of size $n$ and activity $\gap$.\\
$I_n(t)$ & : & Prevalence of the disease within all groups of size $n$.\\
\end{tabular}
\end{center}
\label{tab:TableOfNotation}
\end{table*}

\begin{figure*}[t]
    \centering
    \includegraphics[width=0.8\linewidth]{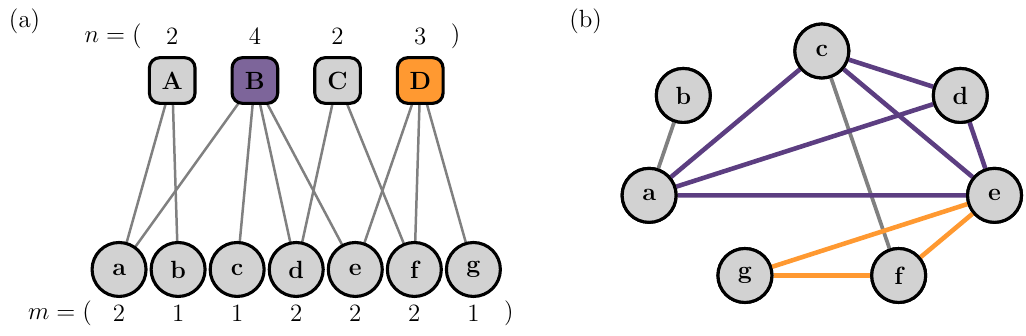}
    \caption{%
      Two representations of a small interaction network composed of 7 individuals (circles) and 4 groups.
      (a) In the bipartite graph representation, an individual $i$ with membership $m_i$ is connected to $m_i$ groups (rounded squares), and a group $j$ of size $n_j$ is connected to $n_j$ individuals.
      The sequences of sizes and memberships are shown above and below the nodes, respectively.
      (b) In the individuals-only projection representation, groups correspond to cliques, meaning that two individuals are connected if and only if they belong to a same group. The links due to the groups of size 3 and 4 are highlighted in orange and purple, respectively, to show the correspondence between both representations.
    }%
    \label{fig:interaction-network}
\end{figure*}

\section{Introduction}
Epidemics are not solely biological phenomena; they are deeply social processes, with disease transmission intricately shaped by human behavior and interaction.
While classical epidemiological models have provided foundational insights into the dynamics of infection spread, they often neglect the particular effects of behavioral responses that significantly influence epidemic trajectories.
Addressing this limitation requires models that account for behavioral heterogeneity and the complex structure of human social networks.

In this study, we focus on four key behavioral and structural factors: (i) the adoption of non-pharmaceutical interventions (NPIs), such as mask-wearing, social distancing, and hand hygiene; (ii) behavioral homophily (the tendency of individuals to interact preferentially with others who share similar behavioral traits); (iii) the organization of interactions into higher-order group structures, such as cliques or communities; and (iv) conformity, the modification of individual behavior in response to group norms or social pressure.

The efficacy of NPIs in mitigating disease spread has been widely documented across multiple outbreaks, where preventive behaviors such as mask-wearing and physical distancing have played critical roles~\cite{talic_effectiveness_2021}, and have also proven essential for surveillance ~\cite{bianconi2021tracing, valganon2024surveillance}.
These behaviors are often associated with behavioral homophily: individuals who adopt preventive measures tend to cluster in the network, forming assortative mixing patterns.
Empirical evidence supports the presence of homophily not only for mask usage~\cite{Haischer2020}, but also for vaccination decisions~\cite{Salath2008} and the adoption of contact-tracing technologies~\cite{Salath2020,rodriguez2021population}.
Models incorporating behavioral homophily have revealed complex, parameter-dependent outcomes, where assortative interactions can either hinder or facilitate epidemic control~\cite{Burgio2021}.

Group structures in social networks serve two important functions.
First, they reflect the mesoscale organization of real-world interactions (within households, workplaces, schools, and social events), where interactions often occur between more than two individuals~\cite{House2008}.
Second, they provide a natural substrate for higher-order interactions, which are essential to capture collective phenomena such as conformity.
Empirical studies of social behavior have demonstrated that social pressure can strongly influence adherence to preventive measures, and that population-level support alone may not predict individual compliance~\cite{Woodcock2021}.
Incorporating conformity into epidemic models has shown that higher-order social influences can qualitatively alter disease dynamics compared to pairwise models~\cite{Zhuang2017, burgio_spreading_2023}.

Motivated by these considerations, we develop a model of the susceptible-infected-susceptible (SIS) process that integrates all four behavioral and structural features outlined above.
It is known that disease, information, and behavior can all localize around certain groups, but the interaction of these phenomena is poorly understood~\cite{Bedson2021}.
In particular, we examine how behavioral homophily interacts with conformity in the context of large group structures.
To analyze these dynamics, we employ the approximate master equations (AME) framework~\cite{HbertDufresne2010, Marceau2010, Gleeson2011, StOngeAME, st-onge_social_2021, Mean_FLAME}, which is well suited to capture localization effects~\cite{StOngeAME} and higher-order interactions in structured populations~\cite{St-Onge2025DefiningClassifyingModels}.
Our results uncover a rich phenomenology, demonstrating that the impact of homophily and conformity depend strongly on group size and composition, with significant implications for epidemic thresholds and containment strategies.

The paper is structured as follows. In section~\ref{model}, we specify the interaction structure considered by the model.
We then set up the behavior of the groups under conformity in section~\ref{group behavior}, before giving the set of approximate master equations describing the dynamics in section~\ref{dynamics}.
Then, in section~\ref{stationary state}, we identify the stationary state, for which we obtain self-consistent equations, and demonstrate the existence of a mesoscopic localization regime.
In section~\ref{homophily}, we examine the impact of homophily on the stationary state and the system's behavior, focusing on its impact on the epidemic threshold in section~\ref{threshold}.
Additionally, in section~\ref{max group size}, we highlight the impact of large groups on the dynamics, showing that both mesoscopic localization and the impact of homophily on the threshold depend on the presence of large groups in the structure.

\section{SIS model with group interactions and individual preferences}
\label{model}
%

\subsection{Structure of interactions}
Our model combines two different types of interactions: epidemic transmission, which is a pairwise process, and behavior modification, which is a group-based, hence higher-order, interaction.
To accurately describe both, we adopt a bipartite graph representation of higher-order structures (see Fig. \ref{fig:interaction-network}), where individual nodes are connected to groups, which represent the subgraphs within which both types of interactions take place~\cite{Newman2003, HbertDufresne2010}.
Functionally, in the individual-only projection representation, these groups are translated into \textit{cliques}, that is, complete subgraphs of $l$ nodes and $\binom{l}{2}$ edges, and each of those edges can transmit the disease.

We refer to the number of groups to which an individual belongs as its \textit{membership}, $m$, and to the number of individuals belonging to a group as its \textit{size}, $n$.
The structure of the network is specified by two distributions: the membership distribution, labeled $g_m$ for $m\in \{m_\mathrm{min}, \dots, m_\mathrm{max}\}$, and the group size distribution, labeled $p_n$ for $n \in \{n_\mathrm{min}, \dots, n_\mathrm{max}\}$.
When the context calls for specific distributions, we consider heterogeneous distributions of the form $p_n \sim n^{-\gamma_n}$ and $g_m \sim m^{-\gamma_m}$.

To represent individual attitudes towards NPIs, we split the population into two types: active ($+$) and passive ($-$).
Although this binary distinction is simplistic, it captures a rich phenomenology and aligns with existing literature~\cite{competing_granell_2014, dynamical_granell_2013, granell2024probabilistic, effect_teslya_2022, effect_fan_2016, effects_kan_2017, impact_pan_2018, effects_pei_2024}.
Active types exhibit a preference for the adoption of NPIs, whereas passive types prefer non-adoption.
The fraction of the population that is active (resp. inactive) is $A_+$ (resp. $A_- = 1 - A_+$).

An important feature of our model is the presence of behavioral homophily, that is, the preference for individuals of the same type to gather in groups.
Social homophily is a well-documented phenomenon~\cite{McPherson2001}, notably in vaccination, where it can greatly influence the epidemic threshold, herd immunity, and final epidemic sizes~\cite{Salath2008, Hiraoka2022}, and in the adoption of digital proximity tracing apps~\cite{Salath2020, Burgio2021}.
Given that in our model, the expressed behavior is determined by group composition, homophily can have a significant impact on disease transmission.

To account for homophily, we define the \textit{composition distribution} for groups, giving the probability that a group of size $n$ contains $n^+$ active individuals, $P_{n^+|n}$. We model this distribution as a mixture of two binomial distributions:
\begin{align}
    P_{n^+|n}
      & = A_+\binom{n}{n^+} \Big( A_+\!+\!hA_- \Big)^{n^+}\!\Big( A_-\!-\!hA_- \Big)^{n-n^+} \nonumber \\
      + \ & A_-\binom{n}{n^+} \Big( A_+\!-\!h A_+ \Big)^{n^+} \Big( A_-\!+h A_+ \Big)^{n-n^+}
    \label{eq:composition}
\end{align}
where $h \in [0, 1]$ is the homophily parameter.
In the absence of homophily ($h=0$), this distribution simplifies to a binomial distribution with parameter $A_+$, meaning that groups are drawn uniformly at random from the population.
When homophily is maximal ($h=1$), only completely homogeneous groups are possible ($P_{n^+=n|n} = A_+$ and $P_{n^+=0|n} = A_-$; all other probabilities are zero).

\subsection{Behavior of the groups}
\label{group behavior}
We now define how individuals within a group adjust their behavior in response to the preferences of other group members.
This response encapsulates the idea of conformity.
Even if an individual has a personal preference for the adoption of NPIs (i.e.
it is active or passive), the presence of other individuals with the opposite preference can affect the expressed behavior of that individual within the group.
For simplicity, we will consider that all individuals within a group adopt the same behavior.
To avoid confusion in the notation, the behavior of the groups is always written between brackets, i.e. $\ga$ and $\gp$.

\begin{figure}
    \centering
    \includegraphics[width=\linewidth]{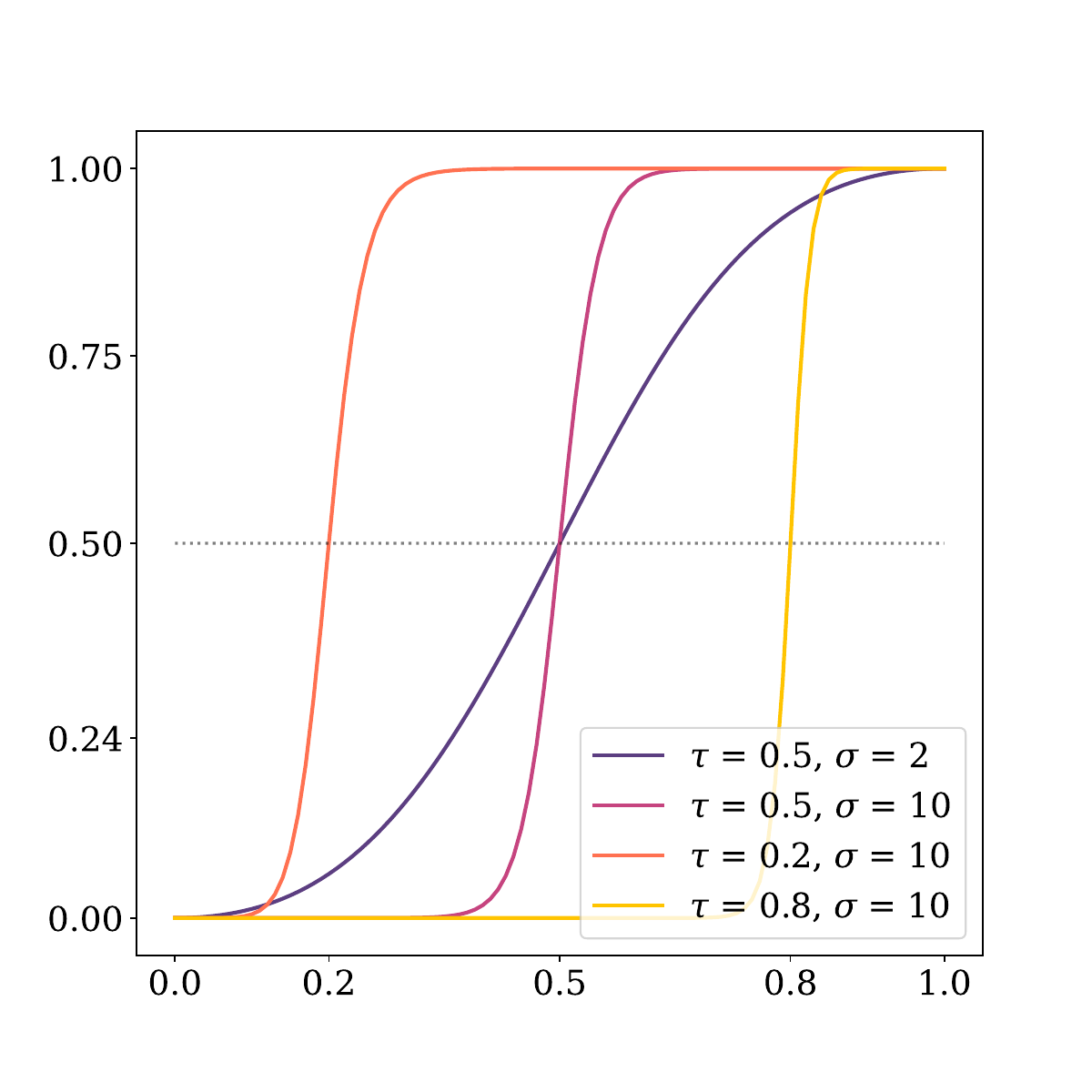}
    \caption{%
      Response probability of Eq.~\ref{eq:response}, giving the probability that a group is active $P_{\ga|n, n^+}$ for a given fraction of active individuals in the group $n^+/n$.
      Each curve represents a pair of influence parameter $\tau$, giving the value of $n^+/n$ for which the probability is exactly $\frac{1}{2}$ --- as shown by the dotted line --- and steepness parameter $\sigma$.%
    }%
    \label{fig:s-curves}
\end{figure}

To fix the behavior of the groups, we define the \textit{response probability} $P_{\ga|n, n^+}$ giving the probability that a group of size $n$ adopts the active behavior given that it contains $n^+$ active individuals.
This probability is arbitrary, but should go from $0$ when there are no active individuals in the group ($n^+=0$), to $1$ when there are only active individuals in the group ($n^+=n$).
For the remainder of this article, we will use the flexible S-curve
\begin{align}
    P_{\ga|n, n^+}
      & = \frac{(\nicefrac{n^+}{n})^{r\sigma}}{\qty(1-(\nicefrac{n^+}{n})^{r})^{\sigma} + (\nicefrac{n^+}{n})^{r\sigma}}, \label{eq:response} \\
      & = 1 - P_{\gp|n, n^+} \nonumber
\end{align}
where $r = \nicefrac{-\ln(2)}{\ln(\tau)}$ with $\tau \in [0, 1]$, and where $\sigma$ controls the steepness of the curve ($\sigma \to \infty$ corresponds to a step function; see Fig.~\ref{fig:s-curves} for examples).
We refer to $\tau$ as the \textit{influence threshold} prescribing the fraction of active individuals ($\nicefrac{n^+}{n}$) for which $P_{\ga|n, n^+}=P_{\gp|n, n^+}$ (i.e., the probability of the group being active is exactly $0.5$).
This parameter reflects how easily groups adopt preventive behaviors: higher values of $\tau$ can be interpreted as measures with greater social cost, requiring a larger active fraction before collective adoption.
On the other hand, lower values of $\tau$ depict situations with low social cost or strong normative pressure, where even a small active minority can trigger group compliance.
We found that the steepness has a very limited impact on the dynamics and is set to $\sigma=10$ henceforth.

\subsection{Dynamical equations}
\label{dynamics}
We now turn to describing the SIS dynamics on this structure of interaction using the approximate master equation framework~\cite{HbertDufresne2010, StOngeAME}.
This framework describes the dynamics exactly within quenched motifs (e.g. groups) that are in turn weakly coupled by some heterogeneous mean-field approximation.
The description is therefore exact in the limit of vanishing coupling.
However, an exact quenched description of groups would here involve tracking their composition based on the type \textit{and} epidemiological state of its members, which yields a number of possible states scaling with the maximum group size as $n_\mathrm{max}^4$.
To simplify the dynamics, we average over possible group compositions, focusing on their type (active or passive) as well as their number of infectious members $i$ and their total number of members $n$.
Similarly, a full description would involve distinguishing individuals by the numbers of active and passive groups to which they belong, but we again simplify the description by simply tracking their own state and type as well as their membership.

Importantly, this approximation not only reduces the complexity of the resulting model by a factor of $n_\mathrm{max}^2$, but allows the model to remain exact in the limit of vanishing coupling where all individuals belong to a single group.
In the limit of disconnected groups, the internal dynamics of groups is completely specified by their type, size, and number of infectious nodes.
The type of the members in a group only serves to induce correlations in the mean-field coupling between groups, which we can approximate at low cost with an appropriate moment closure.

With this in mind, we first track the distribution of the macroscopic states of the individuals, that is, the probability that an individual of membership $m$ and type $\pm$ is susceptible at time $t$, $v_{\s|m, \pm}(t)$.
Second, we track the macroscopic state distribution of the groups, that is, the probability that a group of size $n$ and behavior $\gap$ contains $i$ infected individuals at time $t$, $c_{i| n, \gap} (t)$.
The time evolution of these distributions is given by the following equations:
\begin{subequations}
    \begin{align}
        \dv{v_{\s|m, \pm}}{t}
          & = \mu(1-v_{\s|m, \pm}) - m r^\pm v_{\s|m, \pm} \label{eq:dynamics1} \\
        \dv{c_{i| n, \gap}}{t}
          & = \mu(i+1)c_{i+1|n, \gap} - \mu i c_{i|n, \gap} \nonumber \\
          & \quad + (n-i+1)\qty[\beta^\gap(i-1)+\rho_n^\gap]c_{i-1|n, \gap} \nonumber \\ &\quad - (n-i)\qty[\beta^\gap i + \rho_n^\gap]c_{i| n, \gap}, \label{eq:dynamics2}
    \end{align}
    \label{eq:dynamics}
\end{subequations}
where $\mu$ is the recovery rate, $\beta^\gp$ is the unprotected infection rate, and $\beta^\ga$, the protected infection rate.
For convenience, we will now define $\beta^\gp =: \beta$ as the natural transmission rate, and $\beta^\ga = \epsilon \beta$, with $\epsilon \in [0, 1]$ representing the efficiency of the NPIs at preventing the spread of the infection.
Going forward, we rescale the time and the infection rates by the expected recovery period $1/\mu$, and work with the dimensionless dynamical system without loss of generality.
We will also omit writing the time dependency $(t)$, unless we wish to emphasize the time evolution of a given variable.

In equations \ref{eq:dynamics1} and \ref{eq:dynamics2}, $r^\pm$ and $\rho_n^\pm$ are the mean fields describing the interaction between the groups.
The first, $r^+$ (resp. $r^-$) represents the average infection rate received from a randomly chosen group to which the susceptible active (resp. passive) individual belongs, that is,
\begin{align*}
    r^\pm & = \frac{\sum_{n}p_n\sum_{\gap}P_{\gap|n, \pm}\sum_{i=0}^{n}\beta^\gap i(n-i)c_{i|n, \gap}}{\sum_{n}p_n\sum_{\gap}P_{\gap|n, \pm}\sum_{i=0}^{n}(n-i)c_{i|n, \gap}}.
    \label{eq:r}
\end{align*}
Here, $P_{\ga|n,\pm}$ is the probability that a group is active knowing that it is of size $n$ and that it was reached from an individual of type $\pm$. It is given by (see Appendix A for details)
\begin{align}
    P_{\ga|n, +} &= \frac{\sum_{n^+} n^+ P_{n^+|n}P_{\ga|n, n^+}}{\sum_{n^+} n^+P_{n^+|n}} = 1 - P_{\gp|n, +}\\
    \label{eq:behavior_from_type}
    P_{\ga|n, -} &= \frac{\sum_{n^+} (n-n^+) P_{n^+|n}P_{\ga|n, n^+}}{\sum_{n^+} (n-n^+)P_{n^+|n}} = 1 - P_{\gp|n, -}. \nonumber
\end{align}
In sum, if we choose a group at random from the groups to which a node of type $\pm$ belongs, the joint distribution of the size, behavior and number of infected individuals of that group is proportional to $p_nP_{\gap|n, \pm}(n-i)c_{i|n, \gap}(t)$.
Then, $r^\pm$ is simply the average of the infection rate $\beta^\gap i$ over this distribution.

The second mean field, $\rho_n^\ga$ (resp. $\rho_n^\gp$), gives the average infection rate that a susceptible individual selected at random from an active (resp. passive) group of size $n$ receives through all the \textit{other} groups to which it belongs.
As such,
\begin{align}
    \rho_n^\gap & = \sum_{\pm} A_{\pm|\s, n, \gap} r^\pm\qty[\frac{\sum_m (m-1)mv_{\s|m, \pm}g_m}{\sum_m mv_{\s|m, \pm}g_m}].
\end{align}
The term within brackets gives the mean \textit{excess membership} of susceptible individuals, the distribution of which is proportional to $(m-1)mv_{\s|m, \pm} g_m$.
We then simply multiply this excess membership by the average rate of infection received by the selected individual, $r^\pm$, and by the probability that a susceptible individual drawn at random from a group of size $n$ and behavior $\gap$ is active or passive, $A_{\pm|\s,n,\gap}$.
These probabilities are given by
\begin{align}
    A_{\pm|\s,n,\gap} & =
      \frac{%
        A_{\pm|n,\gap} \sum_m v_{\s|m, \pm} g_m
      }{%
        \sum_{\pm} A_{\pm|n,\gap} \sum_m v_{\s|m, \pm} g_m
      }
    \label{eq:susceptible_sampling}
\end{align}
where $\sum_m v_{\s|m, \pm} g_m$ is the probability that an individual of type $\pm$ is susceptible, and where $A_{\pm|n, \gap}$ is the probability of selecting an individual of type $\pm$ in a group of size $n$ and behavior $\gap$.
These probabilities are given by (see Appendix A)
\begin{align*}
    A_{+|n, \gap}
      & = \frac{%
        \sum_{n^+} (n^+ / n) P_{\gap | n, n^+} P_{n^+ | n}
        }{%
        \sum_{n^+} P_{\gap | n, n^+} P_{n^+ | n}
        } \\
      & = 1 - A_{-|n, \gap}.
\end{align*}

\begin{figure*}[t]
    \centering
    \includegraphics[width=\linewidth]{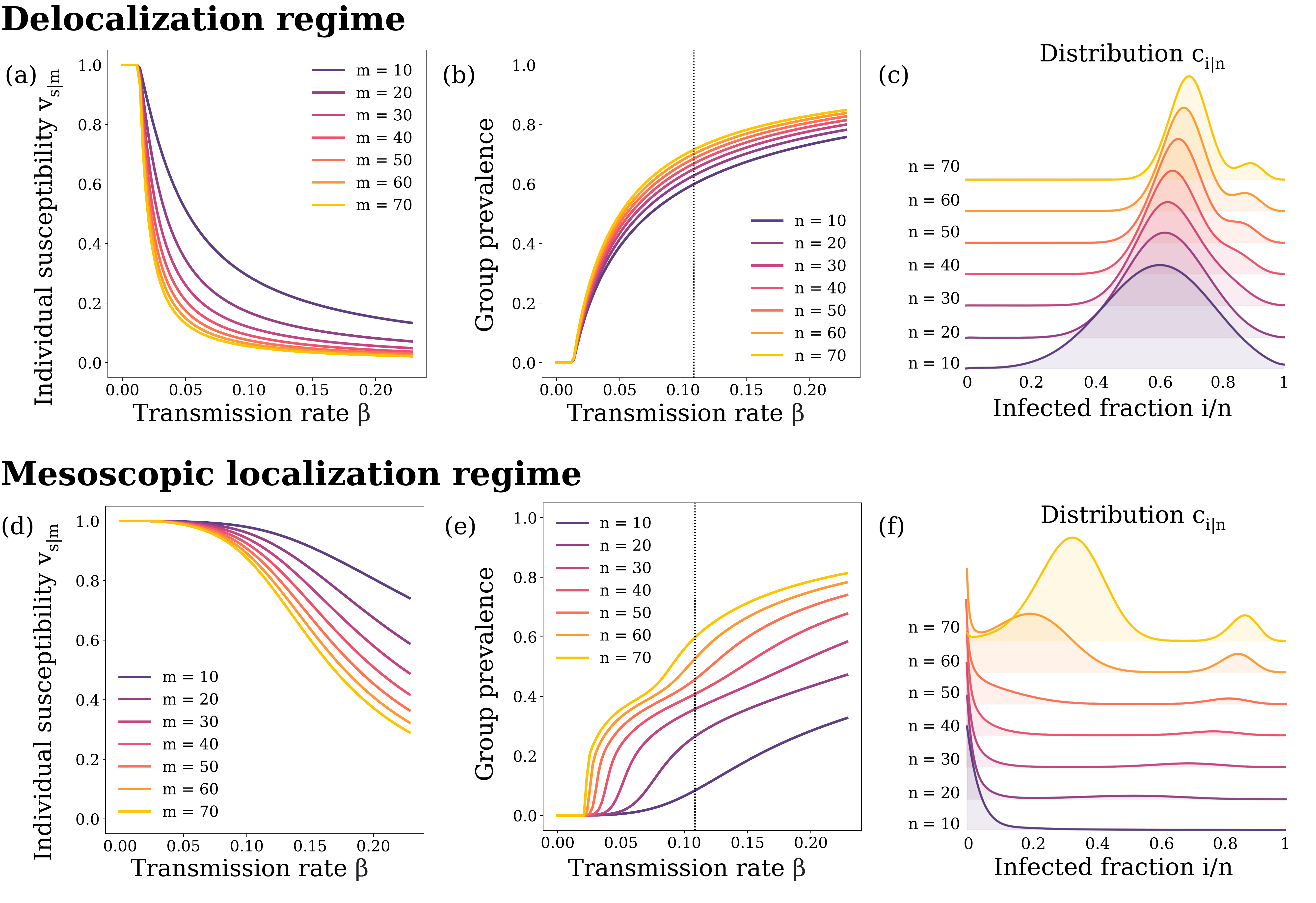}
    \caption{%
      Stationary state as a function of the unprotected transmission rate $\beta$, with heterogeneous distributions $g_m = m^{-\gamma_m}$ and $p_n = n^{-\gamma_n}$, an evenly split population ($A_+=0.5$), influence threshold $\tau = 0.4$, efficiency of NPIs $\epsilon = 0.2$ and no homophily.
      (a)-(c): $\gamma_m = \gamma_n = 2.3$, which yields strong coupling between groups and falls within the delocalized regime.
      (d)-(f): $\gamma_m = 3.3$ and $\gamma_n = 4.0$, which result in weak coupling and a mesoscopic localization regime.
      (a) and (d): individual susceptibilities $v_{\s|m}$ as a function of $\beta$ for different values of membership $m$.
      In the localized regime, the decay is slower because individuals effectively receive infection pressure from fewer groups.
      (b) and (e): group prevalence as a function of $\beta$ for different group sizes $n$.
      In plot (e), corresponding to the localized regime, we observe a double smeared transition: passive groups begin to sustain infection earlier, while active groups follow only later.
      (c) and (f): distribution of the infected fraction across groups of different sizes, evaluated at the value of $\beta$ corresponding to the dotted line in (b) and (e).
      Notice the bimodal distribution, showing higher prevalences in passive groups.
      In (f), notice the mode being $0$ for all group sizes except $n_\mathrm{max} = 70$, demonstrating the localization phenomenon and the absence of transmission within the smaller groups.
    }%
    \label{fig:localization}
\end{figure*}

We can extract various global (or \textit{macroscopic}) observables describing the time evolution of the outbreak from the detailed (or \textit{microscopic}) individual susceptibilities ($v_{\s|m, \pm}$) and group prevalences ($c_{i| n, \gap}$) obtained by integrating Eqs.~\eqref{eq:dynamics} numerically.
First, the probability that an individual of membership $m$ is susceptible, irrespective of its type, is given by
\begin{align}
    v_{\s|m}(t) &= \sum_{\pm} A_{\pm} v_{\s|m, \pm}(t).
\end{align}
Second, the probability that a group of size $n$, irrespective of its behavior, contains $i$ infected individuals is defined as
\begin{align}
    c_{i| n}(t) &= \sum_{\gap} P_{\gap|n} c_{i| n, \gap}(t),
\end{align}
where $P_{\gap|n}$ is the probability that a group of size $n$ has the behavior $\gap$ (see Appendix A)
\begin{align}
    P_{\gap|n} = \sum_{n^+} P_{\gap | n, n^+} P_{n^+ | n}.
\end{align}
Using these distributions, we can now define the prevalence within groups of size $n$ as
\begin{align*}
    I_n(t) &= \sum_i \frac{i}{n}c_{i|n}(t),
\end{align*}
which can be restricted to groups of a given behavior as
\begin{align*}
    I_{n, \gap}(t) &= \sum_i \frac{i}{n}c_{i|n, \gap}(t).
\end{align*}

\subsection{Stationary state}
\label{stationary state}
In the limit $t \to \infty$, the dynamics~\ref{eq:dynamics} reach a fixed point defined by $\dv{v_{\s|m, \pm}}{t} = 0$ $\forall m$ and $\dv{c_{i| n, \gap}}{t} = 0$ $\forall n, i$.
Solving the latter yields the following self-consistent equations for the stationary state:
\begin{subequations}
\begin{align}
    v_{s|m, \pm}^*
      & = \frac{1}{1+mr^\pm}\\
    c_{i| n, \gap}^*
      & = c_{0| n, \gap}^* \binom{n}{i}\prod_{j=0}^{i-1}[\beta^\gap j + \rho_n^\pm].
    \label{eq:stationary_state}
\end{align}
\end{subequations}
While these equations cannot be solved analytically to yield a closed form for the stationary state, they are easily solved numerically through iteration.

\begin{figure*}[t]
    \centering
    \includegraphics[width=0.95\linewidth]{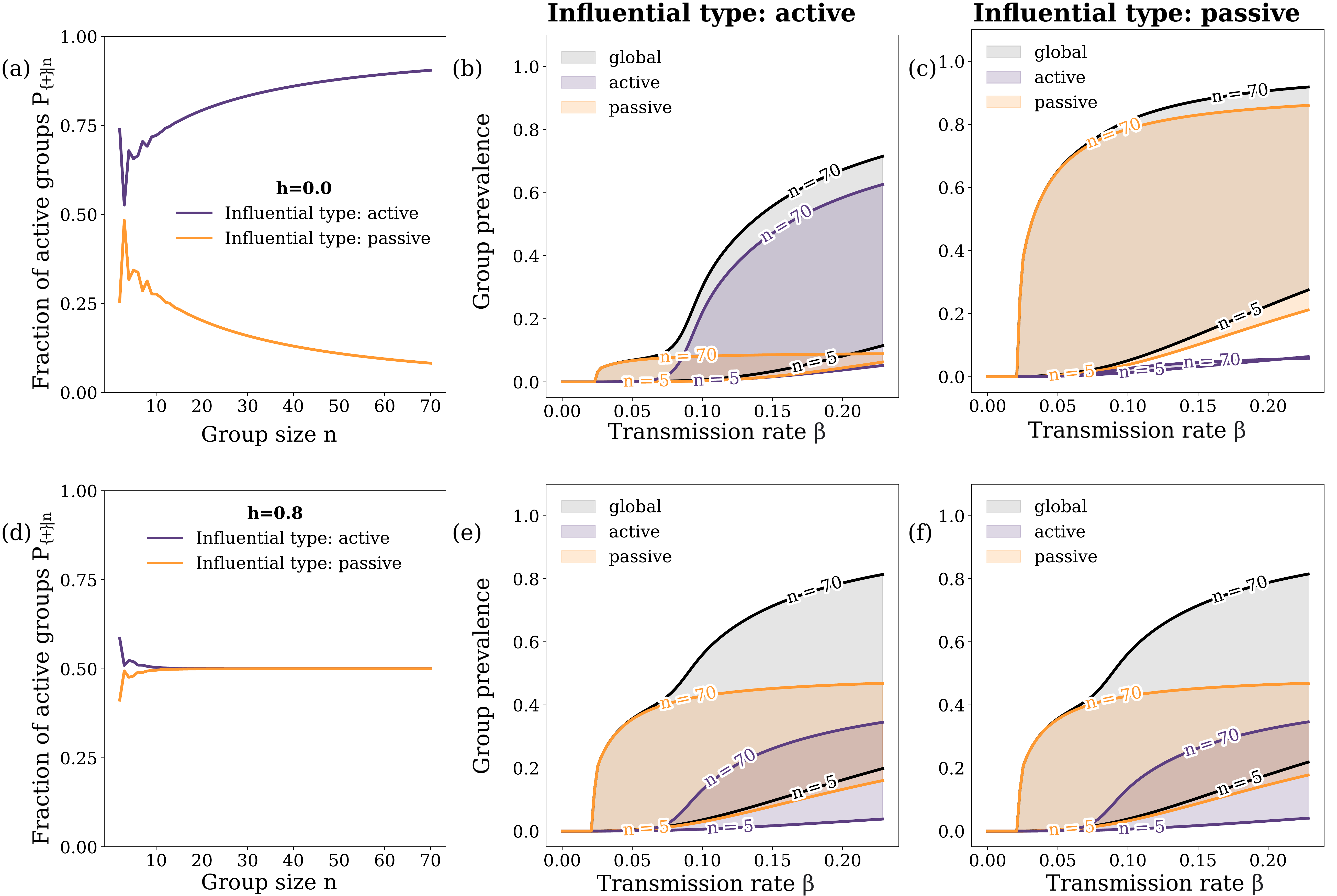}
    \caption{%
      Impact of homophily on the contribution of each group type to the stationary state.
      Here, $\gamma_n = 4.0$ and $\gamma_m = 3.3$, with $n_\mathrm{min}=2$ and $n_\mathrm{max}=70$, such that the dynamics are in the localized regime, and $A_+=0.5$, such that the population is evenly split.
      (a)-(c): No homophily, $h=0$.
      (d)-(f): High homophily, $h=0.8$.
      (a) and (d): expected active fraction of groups as a function of group size $P_{\ga|n}$, with the puple curve corresponding to $\tau=0.4$, such that $A_+>\tau$ and the orange curve, to $\tau=0.6$, such that $A_+<\tau$.
      (b)-(c) and (e)-(f): stationary states represented as the spread between the prevalence in small groups of size $5$ and large groups of size $n_\mathrm{max}$ for active groups, passive groups and both, weighted by the fraction of each group type in the population.
      This way, the correspondence between the global and type-specific dynamics is shown for both small and large groups.
      (b) and (e): $\tau = 0.4$.
      (c) and (f): $\tau = 0.6$.
    }%
    \label{fig:influential-type}
\end{figure*}

As in Ref.~\cite{StOngeAME}, heterogeneous membership distribution $g_m \sim m^{-\gamma_m}$ and group size distribution $p_n \sim n^{-\gamma_n}$, give rise to two distinct stationary–state regimes.
The key quantity determining which regime occurs is the \textit{coupling} between groups, defined as
\begin{equation*}
    \Omega(g_m, p_n) := \qty(\frac{\expval{m(m-1)}}{\expval{m}})\qty(\frac{\expval{n(n-1)}}{\expval{n}}),
\end{equation*}
which corresponds to the average number of \textit{external neighbors} for an individual chosen at random in a group.
Under strong \textit{coupling} ($\Omega \gg n_\mathrm{max}$; see Ref.~\cite{StOngeAME} for details on the mesoscopic localization regimes), the system is delocalized, meaning that all groups, regardless of their size, sustain the spread of the epidemic when $\beta$ is above threshold.
However, under weak coupling ($\Omega \ll n_\mathrm{max}$), the system is localized, with only the largest groups sustaining the epidemic when $\beta$ is only slightly above threshold.
Figure~\ref{fig:localization} displays the difference between the two regimes, recovering the characteristic smeared phase transition of the localized regime reported in Ref.~\cite{StOngeAME}.

In our model, however, the stationary distribution $c_{i|n}^*$ shows a generalized bimodality (see Fig. \ref{fig:localization}(c) and (f)), where the two modes correspond to active and passive groups, prevalence being higher in passive ones.
This source of heterogeneity produces a distinct phenomenon in the localized regime: a \textit{double} smeared transition in which passive groups start to sustain the epidemic first, followed by active groups.
This result combines the effect of heterogeneous group sizes \cite{st-onge_social_2021} with the effect of heterogeneous behavior across groups \cite{stonge2024paradoxes}.
The latter comes in addition to the structural localization and is, in fact, subordinated to it: in the delocalized regime, despite the higher prevalence in passive groups, no smeared transition is observed.

\section{Effect of homophily}
\label{homophily}
In our model, homophily acts exclusively through the group composition distribution (Eq.~\ref{eq:composition}), and not directly on the epidemic prevalence.
As a result, the effect of homophily on the epidemic dynamics must first be understood in terms of how it reshapes group compositions and, in turn, the behavior adopted by those groups.

Under the approximation that the steepness $\sigma \to \infty$, the response probability $P_{\ga|n, n^+}$ is a Heaviside step function with threshold $\tau$.
This allows us to find upper and lower bounds for the fraction of groups that are active, $P_{\ga|n}$, which converge as $n\to\infty$.
Recalling that $\tau$ is the fraction of active individuals within a group such that $P_{\ga|n, n^+} = P_{\gp|n, n^+}$, we find that (see Appendix B)
\begin{align}
    \lim_{n\to\infty} P_{\ga|n} =
      \begin{cases}
        1   & \mbox{ if } h < \frac{A_+-\tau}{A_+} \\
        A_+ & \mbox{ if } h >\frac{A_+-\tau}{A_+}
    \end{cases}
    \label{eq:alpha_limits_act}
\end{align}
if $A_+>\tau$, or
\begin{align}
    \lim_{n\to\infty} P_{\ga|n} =
      \begin{cases}
        0   & \mbox{ if } h < \frac{\tau-A_+}{1-A_+} \\
        A_+ & \mbox{ if } h > \frac{\tau-A_+}{1-A_+}
      \end{cases}
    \label{eq:alpha_limits_pass}
\end{align}
if $A_+<\tau$.
The (upper and lower) bounds converge to these limits at an exponential rate, with the convergence being faster the further $\tau$ is from the parameters of the binomial distributions that make up the composition distribution (i.e. $A_+ + hA_-$ and $A_+-hA_+$; see Eq.~\ref{eq:composition}).
When $\tau$ is equal to one of those values, one of the bounds does not converge to the limit value at all (see Appendix B).

The results of Eqs.~\ref{eq:alpha_limits_act}~and~\ref{eq:alpha_limits_pass} show that the asymptotic behavior of $P_{\ga|n}$ for large groups is controlled by the relative values of $A_+$, $\tau$, and the homophily parameter $h$.
In particular, in the low-homophily regime, large groups adopt the active behavior with probability $1$ if $A_+>\tau$, whereas they adopt it with probability $0$ if $A_+<\tau$.

This separation of regimes points to the existence of an \textit{influential type}, namely the type that dominates group behavior in the absence of homophily.
As such, the active type is influential whenever $A_+>\tau$, and the passive type is influential otherwise.
For example, if $\tau=0.4$ and the population contains $A_+=0.45$ active individuals, the active type is influential (note that it need not constitute an absolute majority).

The key idea conveyed by Eqs.~\ref{eq:alpha_limits_act}~and~\ref{eq:alpha_limits_pass} is that, for low homophily, the behavior of large groups aligns with the influential type, because group compositions concentrate around the population average.
Conversely, for high homophily, group compositions remain segregated even at large sizes, and the behavior of large groups is therefore split into active and passive groups in proportions set by the population composition, leading to the coexistence of distinct collective behaviors.

The first column of Figure~\ref{fig:influential-type} illustrates these two regimes.
In the low-homophily setup (top plot) $P_{\ga|n}$ approaches either 0 or 1 as $n$ increases, showing that large groups are dominated by the influential type.
In contrast, in the high-homophily regime (bottom plot) the curves converge to the same value, the population active fraction $A_+$, indicating that large groups split into active and passive behaviors in proportions set by the population composition.

The two other columns show the impact of the regimes on the global dynamics of the system.
In the low-homophily regime, the global behavior of the system closely matches the behavior of the influential type, even for smaller groups.
Conversely, in the high-homophily regime, the influential type has almost no impact on the global dynamics, with the stationary state being quasi-identical for either influential type.
Plainly, homophily allows the minority type to express its corresponding behavior when it would have been suppressed by the influential type otherwise.

\begin{figure*}[t]
    \centering
    \includegraphics[width=0.8\linewidth]{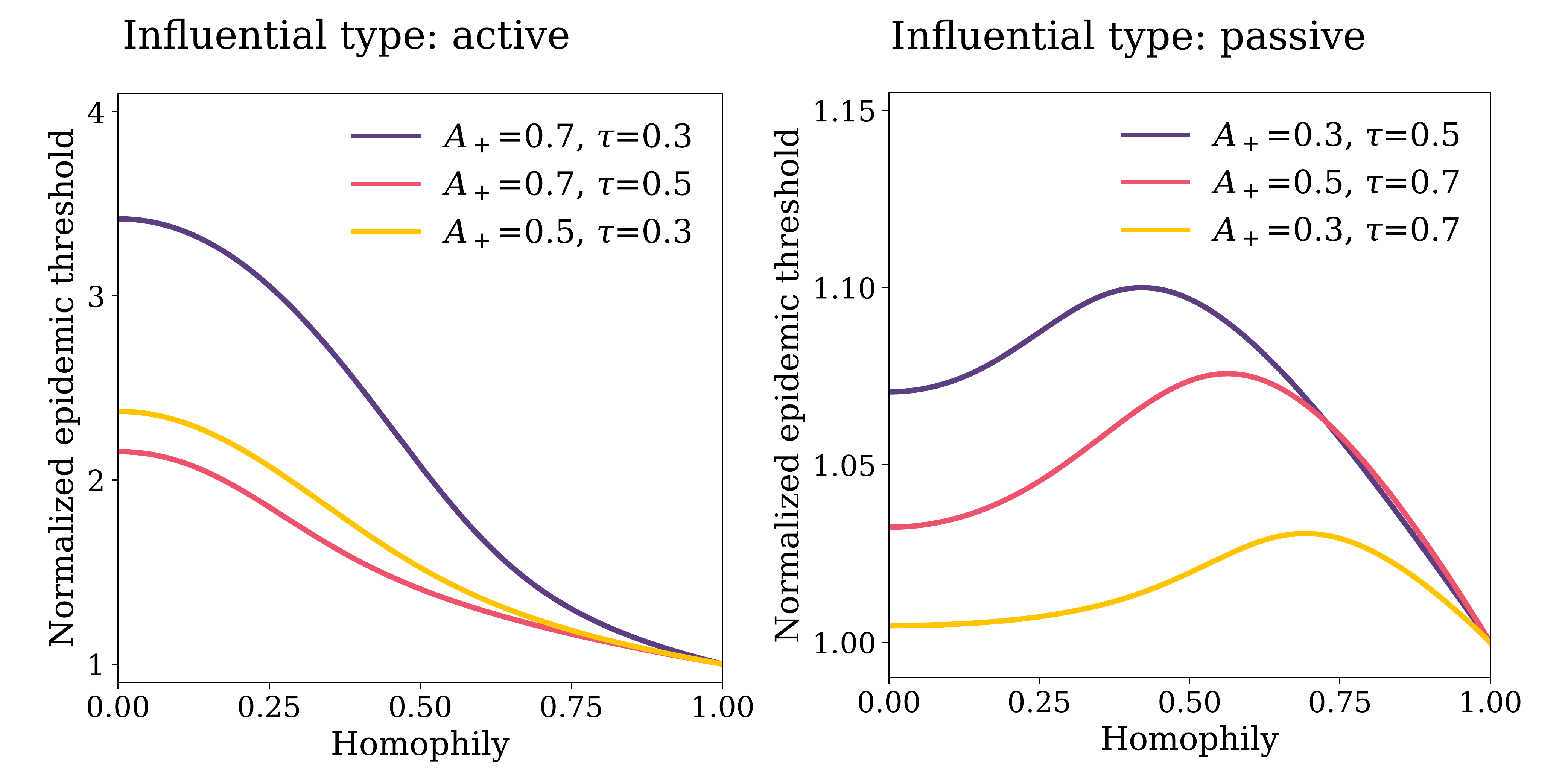}
    \caption{%
      Normalized epidemic threshold as a function of homophily $f(h)$.
      For the whole figure, $\gamma_n = 2.6$, $\gamma_m =3.2$, $n_\mathrm{min} = m_\mathrm{min} = 2$, $n_\mathrm{max}=m_\mathrm{max}=70$ and $\epsilon=0.15$.
      Top: active influential type, $A_+>\tau$. Bottom: passive influential type, $A_+<\tau$.
      The bottom panel displays the non-monotonicity of the threshold as a function of homophily, with the maximum threshold obtained for the value of homophily that maximizes passive individual participation into active groups.
    }%
    \label{fig:threshold-homophily}
\end{figure*}

\subsection{Effect on the epidemic threshold}
\label{threshold}
We now turn to the impact of homophily on the epidemic threshold.
At full homophily ($h=1$), the composition distribution forbids mixed groups, so active and passive individuals never mix in groups.
The result is that the graph splits into two distinct subgraphs: one made up of active individuals and the other made up of passive individuals.
In that scenario, the transmission rate is $\beta$ everywhere on the passive subgraph and $\epsilon\beta$ everywhere on the active subgraph.
Since the two subgraphs share the same membership and group-size distributions, and since $\epsilon<1$ suppresses transmission in the active subgraph, the epidemic threshold is lower in the passive subgraph and therefore determines the global epidemic threshold.
At full homophily ($h=1$), the epidemic threshold $\beta_c(h=1)$ is obtained in Ref.~\cite{StOngeAME} as the solution of a self-consistent equation.
In the limit $n_{\max}\to\infty$, that solution is shown to scale asymptotically as the reciprocal of $\Omega(g_m,p_n)+n_{\max}$, where $\Omega(g_m,p_n)$ is the coupling between groups defined in Section~II.C.
We denote this full-homophily threshold by $\beta_c^*:=\beta_c(h=1)$, and we write the threshold at general homophily as
\begin{align*}
    \beta_c(h)=\beta_c^* f(h).
\end{align*}
An exact closed form for $f(h)$ is not possible to obtain for our model.
However, we can make use of a heuristic to approximate it.
Functionally, when $h<1$, our network is a weighted network, where edges within passive groups have weight $1$ and edges within active groups have weight $\epsilon$.
As such, the effective transmission rate is
\begin{align*}
    \beta_{\mathrm{eff}} = \epsilon \beta w^\ga + \beta (1-w^\ga) = \beta(1-w^\ga + \epsilon w^\ga),
\end{align*}
which yields the epidemic threshold
\begin{align}
    \beta_c(h) = \frac{\beta_c^*}{1 - w^\ga + \epsilon w^\ga}.
    \label{eq:threshold}
\end{align}
Here, $w^\ga$ is the fraction of edges that belong to active groups.
A key nuance, however, is that $w^\ga$ must be calculated on the passive subgraph, that is, the subgraph formed by the passive individuals and their edges (including those connected to active individuals).
The reason for this is that at full homophily ($h=1$), the threshold is determined solely by the passive subgraph, such that $w^\ga =0$ and $f(h) = 1$.
With this in mind, the fraction of active edges in the passive subgraph is
\begin{align*}
    w^\ga &= \frac{\sum_n (n-1) p_n P_{\ga|n,-}}{\sum_n (n-1) p_n} \equiv \frac{\expval{(n-1)P_{\ga|n,-}}}{\expval{(n-1)}},
\end{align*}
where $P_{\ga|n,-}$ is the probability that a group of size $n$ is active, given that it was reached from a passive node, as defined in Eq.~\ref{eq:behavior_from_type}.

Figure \ref{fig:threshold-homophily} shows the dependence of the threshold on homophily, $f(h)$, for different values of the active fraction $A_+$ and of the influence threshold $\tau$.
Importantly, what this figure shows is that when the active type is influential, homophily always decreases the threshold, making the system more susceptible to outbreaks.
By contrast, when the passive type is influential, the effect of homophily is non-monotonic: the threshold is maximized at an intermediate homophily.
The boundary between these two regimes is not exactly $\tau = A_+$.
The value $\tau=A_+$ is therefore only an approximation: finite group sizes induce thresholding effects in $P_{\ga|n}$, so the true transition depends on $\tau$, $A_+$, and the full group-size distribution $p_n$, and does not admit a closed-form expression in general.

The homophily value that maximizes the epidemic threshold is determined by the maximum of $\expval{(n-1)P_{\ga|n,-}}$, that is, the value of homophily at which passive individual participation in active groups is maximized.

The non-monotonic dependence on homophily arises from a  balance between two processes.
Consider first the case when the passive type is influential. Without homophily, groups are dominated by the passive type, per Eqs. \ref{eq:alpha_limits_act} and \ref{eq:alpha_limits_pass}.
Increasing homophily allows more groups to be active, reducing transmission and increasing the epidemic threshold.
However, increasing homophily also suppresses mixing between active and passive types, so the protection associated with active behavior reaches fewer passive individuals, which decreases the threshold.
The optimal homophily is therefore attained when these two effects balance, i.e., when passive individuals are most frequently embedded in active groups.

When the active type is influential, the first effect reverses: increasing homophily only allows more groups to be passive, lowering the threshold. Since both mechanisms act in the same direction, the epidemic threshold decreases monotonically with homophily.

\begin{figure}[b]
    \centering
    \includegraphics[width=\linewidth]{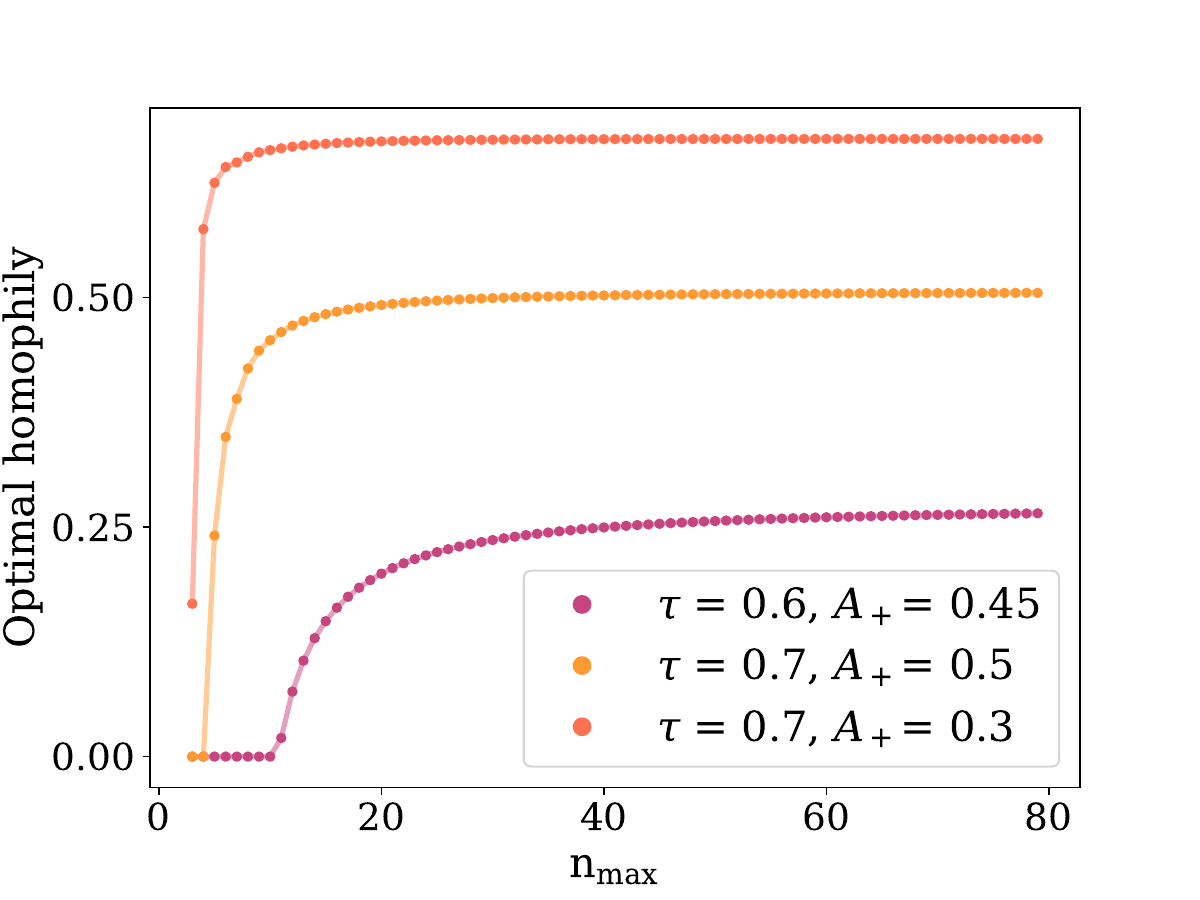}
    \caption{%
    Value of the homophily paramater that maximizes the epidemic threshold as a function of the maximum group size $n_\mathrm{max} \in \{3, ..., 80\}$ for various values of $\tau$ and $A$.
    Here, $\gamma_n$ is tuned for each $n_{\mathrm{max}}$ so that the average degree is constant across all $n_{\mathrm{max}}$ values.
    }%
    \label{fig:impact-n-max}
\end{figure}

\subsection{Effect of maximum group size} \label{max group size}
We now examine how the presence of large groups, controlled by the maximum group size $n_{\max}$, shapes the homophily dependence of the epidemic threshold.
This question is of practical importance because large values of $n_{\max}$ are precisely where most modeling approaches become difficult to apply: the dimension of the dynamical system and the number of required parameters typically grow rapidly with group size, which often forces analyses to focus on small groups.
A central advantage of the approximate master equations (AME) framework is that it can incorporate large group sizes while retaining a tractable and interpretable dynamical description.
By contrast, although many approaches to dynamics on hypergraphs are formally defined for arbitrary $n_{\max}$, they often become impractical at large group sizes.
As a result, numerical studies frequently restrict attention to groups of size $\{2,3\}$~\cite{Iacopini2019, Matamalas2020, BurgioHigher2021, Li2021, Nie2022, CisnerosVelarde2022, Barrat2022, Lucas2023, burgio_spreading_2023, Kim2023, Lin2024}, or to specific uniform-hypergraph settings~\cite{Bodo2016, Jhun2019, Barrat2022}, potentially obscuring higher-order effects driven by larger groups.
In our model, the inclusion of large group sizes is particularly important for behavioral homophily, since large groups are the ones most affected by it.
In particular, because the exact delimitation between the monotonic and non-monotonic regimes for the threshold depends on the full group-size distribution $p_n$, it also depends on its upper cutoff $n_\mathrm{max}$.

Figure \ref{fig:impact-n-max} shows this dependance for various values of $A_+$ and $\tau$.
Crucially, it shows that reducing the maximum group size can not only shift the optimal value of homophily, but can also eliminate it by driving the optimum to $h=0$.
In that case, the threshold becomes monotonic in $h$, indicating that the non-monotonic phenomenology requires sufficiently large groups.

\section{Discussion}

Properly assessing the effect of NPIs is crucial to predict the evolution of epidemics and the potential impact of public health interventions.
Previous work has shown, for instance, that homophily in vaccination can substantially increase the coverage required to achieve herd immunity~\cite{Salath2008}.
Similarly, our results suggest that behavioral homophily can attenuate the population-level effectiveness of widespread mask use by concentrating protection within adoption clusters and leaving other parts of the contact structure comparatively unprotected.
Consequently, conclusions that rely on homogeneous or weakly correlated adoption may underestimate the level of uptake required to strongly suppress transmission.
This point is relevant when interpreting studies suggesting that widespread mask adoption alone could, under certain assumptions, have dramatically reduced COVID-19 transmission in specific settings such as Italy and Brazil~\cite{Gondim2021}: our results suggest that the required coverage may be higher once behavioral clustering is taken into account.

Additionally, our analysis reveals a rich phenomenology arising from the interplay between homophily and conformity, controlled by the popularity of NPIs (through $A_+$), their ease of adoption or social cost (through $\tau$), and the group-size distribution $p_n$.
Most notably, the epidemic threshold exhibits non-monotonic behavior when the NPIs are favored by a minority of the population.
This contrasts with previous studies on behavioral homophily for mask wearing, which showed a monotonically decreasing threshold with homophily~\cite{Watanabe2022}, illustrating the impact of conformity on the dynamics.

Relatedly, behavioral homophily was also shown to have a non-monotonic effect on the epidemic threshold when studying the impact of digital proximity tracing (DPT) apps on epidemic spreading~\cite{Burgio2021}.
Although the DPT setting differs substantially from the present model, the underlying mechanism is closely analogous: decreasing homophily increases the extent to which adopters offer protection to nonadopters, while simultaneously reducing the internal protection within adoption clusters.
The optimal value of homophily balances these opposing effects.
Taken together, these results suggest that non-monotonic threshold responses to homophily may arise whenever an intervention provides protection that is not purely individual, but depends on the participation of neighbors.
In DPT, this neighbor-dependence is explicit by design, whereas in our setting it emerges endogenously through conformity within groups.

Moreover, our results show that the rich phenomenology of the epidemic threshold emerges only when sufficiently large groups are taken into account.
This is especially true when the fraction of active individuals, $A_+$, is close to the adoption threshold in groups, $\tau$.
Mesoscopic localization is also only observed when heterogeneity in group sizes is taken into account \cite{st-onge_social_2021, StOngeAME}.
Large group sizes are therefore important both to accurately determine the vulnerability of the population to epidemic outbreaks, but also to devise public health interventions.
On that front, our model suggests that these interventions should be adapted to the popularity of the NPIs and the cost of their adoption.
For widespread, low-cost NPIs, in addition to targeting the largest groups, the negative impact of homophily suggests that public health interventions should be focused on the clusters of passive individuals, which form the core of the epidemic transmission networks.
For unpopular or costly NPIs, the non-monotonicity of the threshold suggests that quantifying the level of behavioral homophily is important to inform intervention strategies.
If homophily is low, increasing homophily can be beneficial and could be achieved by targeting interventions at the individuals already most likely to adopt NPIs.
Similarly, if homophily is high, targeting the individuals least likely to adopt NPIs can help reduce homophily, and thus help extend protection to more of the population.

Finally, an important limitation of our work is the binary nature of conformity, where all individuals in a group adopt the same behavior.
Related to this limitation is the binary representation of individual attitudes towards NPI adoption.
A more realistic perspective would be to allow each individual to have a certain (continous) awareness value, representing how willing that individual is to adopt NPIs.
To refine the model and more realistically capture individual attitudes and conformity, future work could allow each individual to adopt NPIs with a probability that depends both on their own propensity and on the composition of their group.
We expect the present framework to provide a natural basis for developing and analyzing such extensions.






\begin{acknowledgments}
  This work was supported by the Natural Sciences and Engineering Research Council of Canada (OR, AAl) and the Sentinelle Nord program of Universit\'e Laval, funded by the Canada First Research Excellence Fund (OR, AAl).
  L.H.-D. acknowledges financial support from the National Science Foundation award \#2419733.
  AAr and CG acknowledge support from the Spanish Ministerio de Ciencia e Innovaci\'on (PID2024-158120NB-C21), Generalitat de Catalunya (2021SGR-00633), Universitat Rovira i Virgili (2023PFR-URV-00633), ICREA Academia, and the Joint Appointment Program at Pacific Northwest National Laboratory (PNNL).
  PNNL is a multi-program national laboratory operated for the U.S.\ Department of Energy (DoE) by Battelle Memorial Institute under Contract No.\ DE-AC05-76RL01830.
\end{acknowledgments}

\appendix

\section*{Appendix A. Structural probabilities}

The only free probabilities for the model are the composition distribution \ref{eq:composition} and response probability \ref{eq:response}.
From these two probability distributions, all other probabilities relevant to the model can be derived.
First, let's define the inverse composition distribution $P_{n^+|n, \ga}$, that is, the probability that a group of size $n$ contains $n^+$ active individuals given that the group is active. By the extended form of Bayes' theorem,
\begin{align*}
    P_{n^+|n, \ga} &= \frac{P_{\ga|n, n^+}P_{n^+|n}}{\sum_{n^+=0}^n P_{\ga|n, n^+}P_{n^+|n}}.
\end{align*}
The denominator above is simply the total probability that a group of size $n$ is active, which yields
\begin{align*}
    P_{\ga|n} &= \sum_{n^+=0}^n P_{\ga|n, n^+}P_{n^+|n} = 1 - P_{\gp|n}.
\end{align*}
Similarly, we find that
\begin{align*}
    P_{n^+|n, \gp} &= \frac{P_{\gp|n, n^+}P_{n^+|n}}{\sum_{n^+=0}^n P_{\gp|n, n^+}P_{n^+|n}}.
\end{align*}
Now, we want the same probability but conditioned on the group being reached through an active or passive individual, $P_{\ga|n, +}$ or $P_{\ga|n, -}$.
This introduces a selection bias equal to the number of individuals of the correct type in the group, $n^+$ or $n^- = n-n^+$, as a group is more likely to be reached through an individual of a certain type if it contains more individuals of that type.
As such, for every number of same type individuals, $n^\pm$, this probability is proportional to the product of that number, the composition distribution evaluated at that value, $P_{n^+|n}$, and the response distribution evaluated at that composition, $P_{\ga|n, n^+}$.
Summing over all possible group compositions and normalizing yields the desired probabilities:
\begin{align*}
    P_{\ga|n, +} &= \frac{\sum_{n^+} n^+ P_{n^+|n}P_{\ga|n, n^+}}{\sum_{n^+} n^+P_{n^+|n}} = 1 - P_{\gp|n, +}\\
    P_{\ga|n, -} &= \frac{\sum_{n^+} (n-n^+) P_{n^+|n}P_{\ga|n, n^+}}{\sum_{n^+} (n-n^+)P_{n^+|n}} = 1 - P_{\gp|n, -}.
\end{align*}
Finally, we need the probability of drawing an individual of a given type conditioned on the size and behavior of that group, $A_{\pm | n, \gap}$.
Setting the probability of drawing one of the $n^+$ active individuals in a group of size $n$ as
\begin{align*}
    A_{+|n, n^+} = \frac{n^+}{n} = 1 - A_{-|n, n^+},
\end{align*}
we directly obtain
\begin{align*}
    A_{+|n, \gap}
      & = \sum_{n^+} A_{+|n, n^+} P_{n^+|n, \gap} \\
      & = \frac{\sum_{n^+ = 0}^n (n^+ / n) P_{\gap|n, n^+} P_{n^+|n}}{\sum_{n^+ = 0}^n P_{\gap|n, n^+} P_{n^+|n}} \\
      & = 1 - A_{-|n, \gap}.
\end{align*}

\section*{Appendix B. Homophily regimes}
In the limit $\sigma \to \infty$ for the response probability, the latter becomes a Heaviside step function, and $P_{\ga|n}$ is
\begin{align*}
    P_{\ga|n}
      & = A_+F\qty((1-\tau)n; n, 1-h-A_+(1-h))\\
      & \quad + (1-A_+)F\qty((1-\tau)n; n, 1-(1-h)A_+),
\end{align*}
where $F(k; n, p)$ is the cumulative distribution function of the binomial distribution. While there is no closed form for this function, it is bounded by the following Chernoff bounds, provided that $k<np$:
\begin{align*}
    F(k; n, p) \leq e^{-nD_{KL}\qty(\frac{k}{n}||p)}\\
    F(k; n, p) \geq \frac{1}{\sqrt{2n}}e^{-nD_{KL}\qty(\frac{k}{n}||p)},
\end{align*}
where $D_{KL}\qty(p||q)$ is the Kullback-Leibler divergence between the Bernoulli($p$) and Bernoulli($q$) distributions, given by
\begin{align*}
    D_{KL}\qty(p||q) & = p\ln\frac{p}{q} + (1-p)\ln\frac{1-p}{1-q}.
\end{align*}
If $k>np$, it suffices to use
\begin{align*}
    F(k; n, p) \geq 1-e^{-nD_{KL}\qty(\frac{k}{n}||p)}\\
    F(k; n, p) \leq 1-\frac{1}{\sqrt{2n}}e^{-nD_{KL}\qty(\frac{k}{n}||p)}.
\end{align*}
Therefore, for $P_{\ga|n}$, if $\tau > h+(1-h)A$, then the condition $k<np$ is respected for both functions $F(k; n, p)$ in $P_{\ga|n}$, and using $D_{KL}(p||q) = D_{KL}(1-p||1-q)$,
\begin{align*}
    P_{\ga|n}
      & \leq A_+e^{-nD_{KL}\qty(\tau||h+(1-h)A_+)} \\
      & \quad + A_-e^{-nD_{KL}\qty(\tau||(1-h)A_+)}\\
    P_{\ga|n}
      & \geq \frac{A_+}{\sqrt{2n}}e^{-nD_{KL}\qty(\tau||h+(1-h)A_+)} \\
      & \quad + \frac{A_-}{\sqrt{2n}}e^{-nD_{KL}\qty(\tau||(1-h)A_+)}.
\end{align*}
Similarly, if $\tau < (1-h)A_+$, then $k>np$ for both functions, such that
\begin{align*}
    P_{\ga|n}
      & \leq 1-A_+\frac{1}{\sqrt{2n}}e^{-nD_{KL}\qty(\tau||h+(1-h)A_+)}\\
      & \quad -A_-\frac{1}{\sqrt{2n}}e^{-nD_{KL}\qty(\tau||(1-h)A_+)}\\
    P_{\ga|n}
      & \geq 1-A_+e^{-nD_{KL}\qty(\tau||h+(1-h)A_+)})\\
      & \quad - A_-e^{-nD_{KL}\qty(\tau||(1-h)A_+)}.
\end{align*}
Finally, if $(1-h)A_+ < \tau < h+(1-h)A_+$,
\begin{align*}
    P_{\ga|n}
      & \leq A_+\qty(1-\frac{1}{\sqrt{2n}}e^{-nD_{KL}\qty(\tau||h+(1-h)A_+)})\\
      & \quad + A_-e^{-nD_{KL}\qty(\tau||(1-h)A_+)}\\
    P_{\ga|n}
      & \geq A_+\qty(1-e^{-nD_{KL}\qty(\tau||h+(1-h)A_+)})\\
      & \quad + \frac{A_-}{\sqrt{2n}}e^{-nD_{KL}\qty(\tau||(1-h)A_+)}.
\end{align*}
It is easily seen that in each of these cases, both exponentials go to zero as $n \to \infty$ – as long as $D_{KL}$ is non-zero – such that, by the squeeze theorem, the limit for $P_{\ga|n}$ is $0$ in the first case, $1$ in the second and $A_+$ in the third.
Rewriting the conditions for $\tau$, $A_\pm$ and $h$ for each of the cases in terms of $h$ yields
\begin{align*}
    \lim_{n\to\infty} P_{\ga|n}
      = \begin{cases}
          1   &\mbox{ if } h \in \big[0, \frac{A_+-\tau}{A_+}\big). \\
          A_+ &\mbox{ if } h \in \big(\max \left\{\frac{\tau-A_+}{A_-}, \frac{A_+-\tau}{A_+}\right\}, 1\big] \\
          0   &\mbox{ if } h \in \big[0, \frac{\tau-A_+}{A_-}\big)
        \end{cases}
\end{align*}

\begin{figure*}[t]
    \centering
    \includegraphics[width=\linewidth]{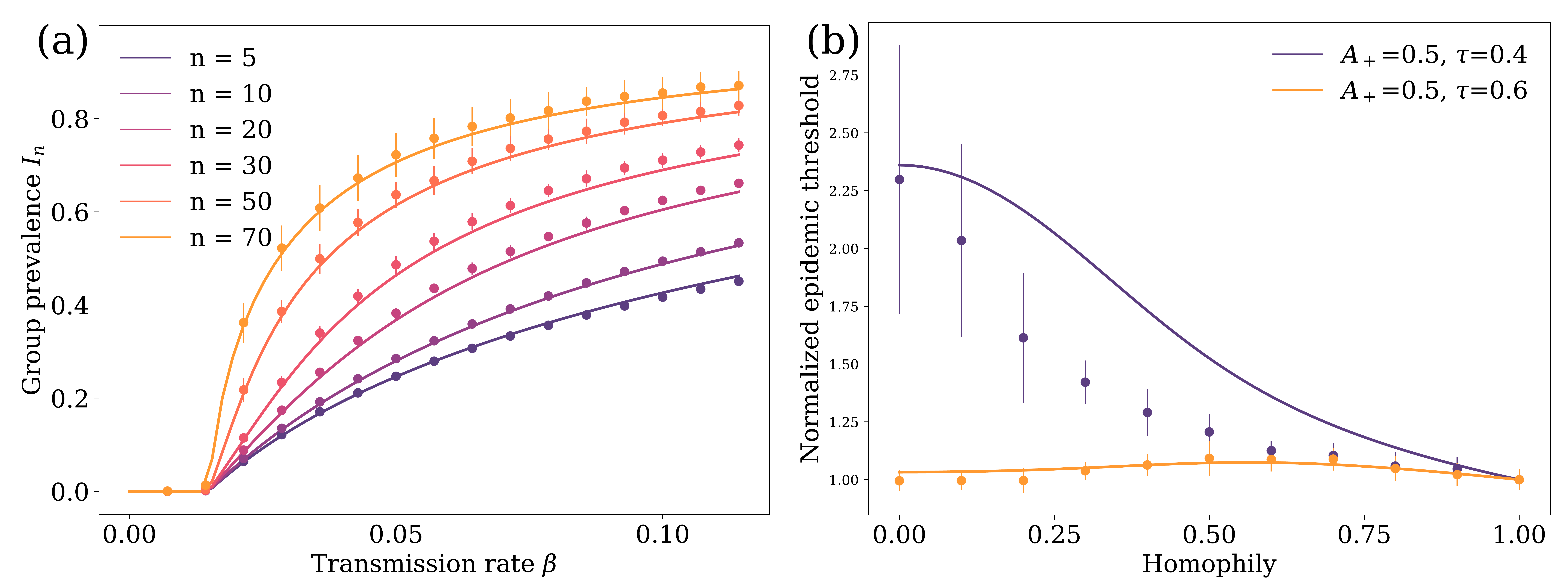}
    \caption{%
      Comparison of the model predictions with Monte-Carlo simulations with parameters $A_+=0.5$, $\epsilon=0.2$, $\gamma_n = 2.6$, $\gamma_m = 3.3$ and $n_\mathrm{max} = m_\mathrm{max} = 70$.
      (a) Comparison for the stationary state, with $\tau = 0.6$ and $h=0.0$.
      The solid lines represent the stationary prevalences as a function of the transmission rate $\beta$, for different group sizes obtained from equations \ref{eq:stationary_state}.
      Each marker represents the average of 800 samples of the stationary state for 50 realization of the network with $N=2\times 10^5$ individuals.
      Error bars (sometimes smaller than the marker) represent one standard deviation.
      The samples were taken after a burn-in period of $2\times 10^6$ events, and were separated by a uniform random decorrelation period between 10 and 50 events.
      (b) Comparison for the dependence of the epidemic threshold on homophily with Monte-Carlo simulations.
      The solid lines represent equation \ref{eq:threshold}.
      Each point represents the average value for the normalized threshold obtained by finding the maximum susceptibility for samples of 1000 states simulated on 50 different networks of $5\times 10^5$ individuals.
      Error bars represent one standard deviation.
    }%
    \label{fig:mc-validation}
\end{figure*}

The rate of convergence is bounded by the slower of the two exponentials, with the slowest being the one for which $D_{KL}$ is the smallest, yielding
\begin{align*}
    \abs{P_{\ga|n}^{*} - \lim_{n\to\infty} P_{\ga|n}}
      \sim \begin{cases}
             e^{-n D_{KL}\qty(\tau || h+(1-h)A_+)} \\ \mbox{ if }\qty(1+\frac{h}{(1-h)A_+})^{\tau}\\ \quad> \qty(1-\frac{h}{1-(1-h)A_+})^{\tau-1}\\
             e^{-n D_{KL}\qty(\tau || (1-h)A_+)} \\ \mbox{ if }\qty(1+\frac{h}{(1-h)A_+})^{\tau}\\ \quad< \qty(1-\frac{h}{1-(1-h)A_+})^{\tau-1}.
           \end{cases}
\end{align*}
When $\tau \in \{(1-h)A_+, h+(1-h)A_+\}$, one of the exponentials is constant, so that the upper and lower bounds don't converge to the same value, and that the limit for $P_{\ga|n}$ is undefined.

\section*{Appendix C. Generating homophilic networks with groups}
To generate graphs with the structure corresponding to the model, we use a stub matching process.
We first assign a membership to each individual with probability corresponding to the membership distribution $g_m$, creating an individual stub list containing each individual as many times as its membership.
We then split these individuals into the active and passive types with a fraction $A_+$ of the individuals being active.

Next, we generate the group stubs under the same principle, with the group size being drawn from the group size distribution $p_n$, and the active fraction of those $n$ stubs drawn from the composition distribution $P_{n^+|n}$.
Since there needs to be as many individual stubs as group stubs, in practice, we generate group stubs until there are at least as many group stubs as individual stubs.
If there are too many group stubs, we remove a group at random.
We repeat the procedure until the number of stubs matches.
Then, to make sure the number of stubs of the same activity match as well, we transfer the necessary amount of group stubs from one activity to the other.
This introduces a deviation from the composition distribution, but in practice, this deviation becomes vanishingly small for large network sizes.

Finally, the individual and group stubs of same activity are matched uniformly at random, creating an edge between all individuals belonging to the same groups.
This allows for multi-edges and self-loops, although in practice, the probability of occurence for these becomes vanishingly small in the large network limit, making their impact on the dynamics negligible.

\section*{Appendix D. Comparison with Monte-Carlo simulations}
To validate our findings, we simulated SIS dynamics on homophilic networks with large cliques.
The results, shown in figure \ref{fig:mc-validation}(a), show that our model is accurate in its prediction of the stationary state.
We also independently tested our theoretical predictions for the dependence of the epidemic threshold on homophily, with the results shown in figure \ref{fig:mc-validation}(b).
Once again, our predictions are validated, including the non-monotonic behavior when the passive type is influential.

To sample the quasi-stationary state, we used the efficient composition and rejection algorithm presented in \cite{StOnge2019}.
For low values of $\beta$, the SIS process on finite networks usually reaches the absorbing state.
To avoid this issue, we implement the hub reactivation procedure described in \cite{Sander2016}, which was shown to accurately capture localized states.
In this procedure, if the system reaches the absorbing state, the epidemic is restarted by infecting one of the individuals with the highest \textit{transmission degree} $k = k^- + \epsilon k^+$.
Allowing the system a new burn-in period after reinfection leads to sampling the quasi-stationary state.

To sample the epidemic threshold, we sampled the quasi-stationary state for various values of $\beta$, calculating the susceptibility $\chi = \frac{\expval{\rho^2}-\expval{\rho}^2}{\expval{\rho}}$ for each of the samples, and identifying the threshold as the value of $\beta$ that maximizes the susceptibility \cite{Ferreira2012}.
While susceptibility maximization is the most accurate method available to us to identify the epidemic threshold, note that it is not exact due to finite size effects.
These, notably, result in a small sample of $n_{\mathrm{max}}$ size groups, leading to deviations from the expected behavior for these groups.
As the largest groups are disproportionally important when it comes to the value of the epidemic threshold, the inaccuracy introduced by their scarcity is particularly large when identifying the latter.

Part of the difference between the Monte Carlo results and our model's prediction in Fig.~\ref{fig:mc-validation} can also be attributed to dynamical correlations within the groups which our model doesn't capture.
Indeed, for the steady-state, when calculating the group infection mean field $\rho_n^\pm$, we make use of the probability of drawing an active individual from the pool of susceptible individuals in the group.
To give this probability exactly, we would need to keep track of the number of infected individuals of each type in groups of size $n$, through dynamical variables such as $c_{i, i^+|n, n^+, \gap}$ instead of the $c_{i|n, \gap}$ we use in our model.
With our dynamical variables, we instead need to replace the exact value for the sampling probability with the population average, $A_{\pm|s, n, \gap}$ (Eq.~\ref{eq:susceptible_sampling}).
While the more detailed model would likely be more accurate, it would be of dimension $\mathcal{O}(n_\mathrm{max}^4)$ (compared to $\mathcal{O}(n_\mathrm{max}^2)$ for our model), greatly hindering the analysis of large group sizes, which our results have shown to be crucial to the dynamics of the system.
As such, we believe our model to be an excellent compromise between interpretability and accuracy.

\end{document}